\documentclass[aps,prb,twocolumn,superscriptaddress,floatfix]{revtex4-1}

\usepackage[percent]{overpic}
\usepackage{graphicx,graphics}
\usepackage{dcolumn}
\usepackage{amsmath,amssymb,amsfonts}
\usepackage{latexsym,verbatim}
\usepackage{bm}
\usepackage{mathtools}
\usepackage{bbold}
\usepackage{color}
\usepackage{ulem}
\usepackage[breaklinks=false,colorlinks,citecolor=blue,linkcolor=blue,urlcolor=blue]{hyperref}

\usepackage{braket}
\usepackage{soul}
\begin{document}
\title{Charger-mediated energy transfer in exactly-solvable models for quantum batteries}

\author{Gian Marcello Andolina}
\email{gian.andolina@sns.it}
\affiliation{Istituto Italiano di Tecnologia, Graphene Labs, Via Morego 30, I-16163 Genova, Italy}
\affiliation{NEST, Scuola Normale Superiore, I-56126 Pisa, Italy}

\author{Donato Farina}
\affiliation{Istituto Italiano di Tecnologia, Graphene Labs, Via Morego 30, I-16163 Genova, Italy}
\affiliation{NEST, Scuola Normale Superiore, I-56126 Pisa, Italy}

\author{Andrea Mari}
\affiliation{NEST, Scuola Normale Superiore, I-56126 Pisa, Italy}

\author{Vittorio Pellegrini}
\affiliation{Istituto Italiano di Tecnologia, Graphene Labs, Via Morego 30, I-16163 Genova, Italy}

\author{Vittorio Giovannetti}
\affiliation{NEST, Scuola Normale Superiore, I-56126 Pisa, Italy}

\author{Marco Polini}
\affiliation{Istituto Italiano di Tecnologia, Graphene Labs, Via Morego 30, I-16163 Genova, Italy}

\date{\today}

\begin{abstract}
We present a systematic analysis and classification of several  models of quantum batteries involving different combinations of two-level systems and quantum harmonic oscillators. In particular, we study energy-transfer processes from a given quantum system, termed ``charger'', to another one, i.e.~the proper ``battery''. In this setting, 
we analyze different figures of merit, including the charging time, the maximum energy transfer, and the average charging power. The role of coupling Hamiltonians which do not preserve the number of local excitations in the charger-battery system is clarified by properly accounting them in the 
global energy balance of the model.
\end{abstract}
\maketitle

\section{Introduction}

Currently there is worldwide interest in exploiting quantum phenomena such as superposition, quantum coherence, and entanglement for future technologies~\cite{Riedel2017,Acin17} in the realms of communication, computation, simulation, and sensing/metrology. On a seemingly disconnetted path, the possibility to use quantum resources to achieve superior performances in the manipulation of energy is currently being intensively studied~\cite{Vinjanampathy16,Alicki18,Strasberg16,Goold16,Campisi11,Horodecki13,Gelbwaser-Klimovsky15,Watanabe17}. 

\begin{figure}[th!]
\centering
\begin{overpic}[width=0.9\columnwidth]{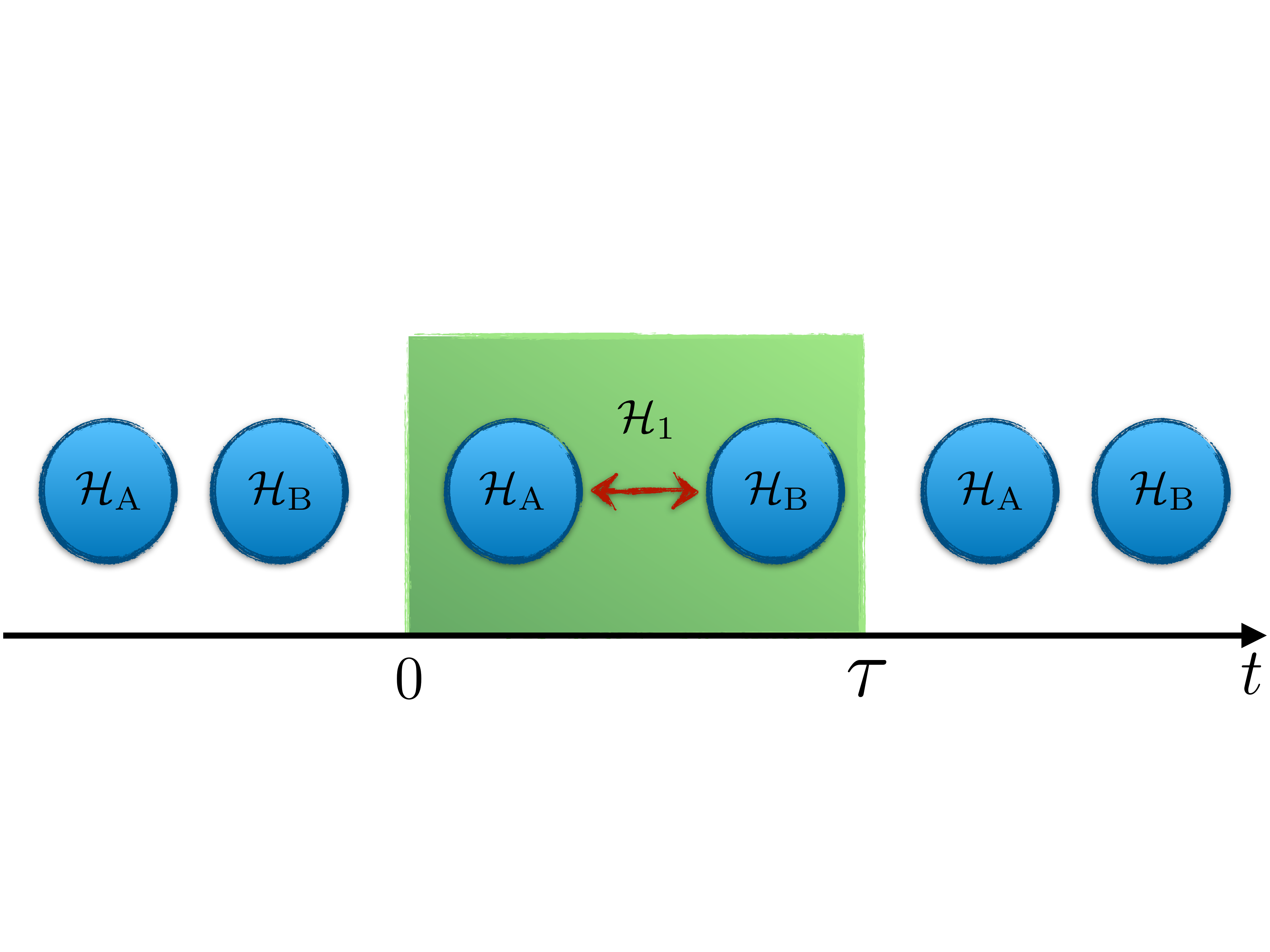} \put(-4,55){\normalsize (a)}
\end{overpic}
\begin{overpic}[width=0.8\columnwidth]{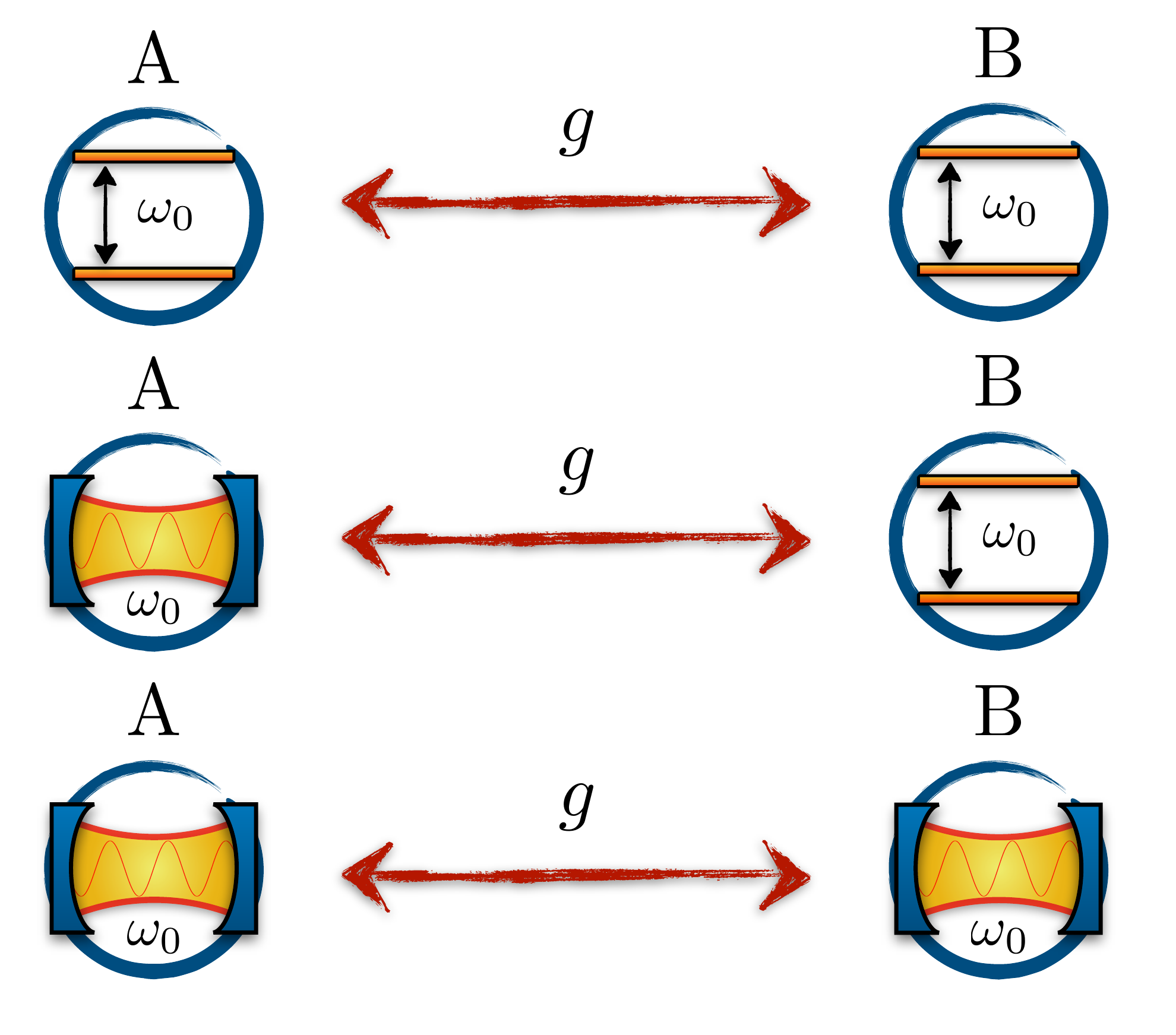}\put(-10,88){\normalsize (b)} \put(-6,78){\normalsize (1)}\put(-6,47){\normalsize (2)}\put(-6,14){\normalsize (3)}\end{overpic}
\caption{(Color online) Panel (a) shows the time-dependent interaction protocol that allows energy flow between the charger, described by the Hamiltonian ${\cal H}_{\rm A}$, and the battery, described by the Hamiltonian ${\cal H}_{\rm B}$. At time $t<\tau$ the two systems A and B do not interact and cannot exchange energy, their dynamics being governed by the Hamiltonian $\mathcal{H}_0=\mathcal{H}_{\rm A}+\mathcal{H}_{\rm B}$. In the time interval $0<t<\tau$ the Hamiltonian $\mathcal{H}_1$ is switched on and the two systems interact. Finally, the interaction is switched off at time $\tau$, and the energy $E_{\rm B} (\tau)$ stored in the battery B  is a conserved quantity.  Panel (b) illustrates cartoons of the three charger-battery toy models introduced and studied in this Article. Sub-panel (1): energy transfer is studied between two qubits; sub-panel (2): energy transfer is studied between a quantum harmonic oscillator and a qubit; sub-panel (3): energy transfer is studied between two quantum harmonic oscillators.  \label{fig:System}}
\end{figure}
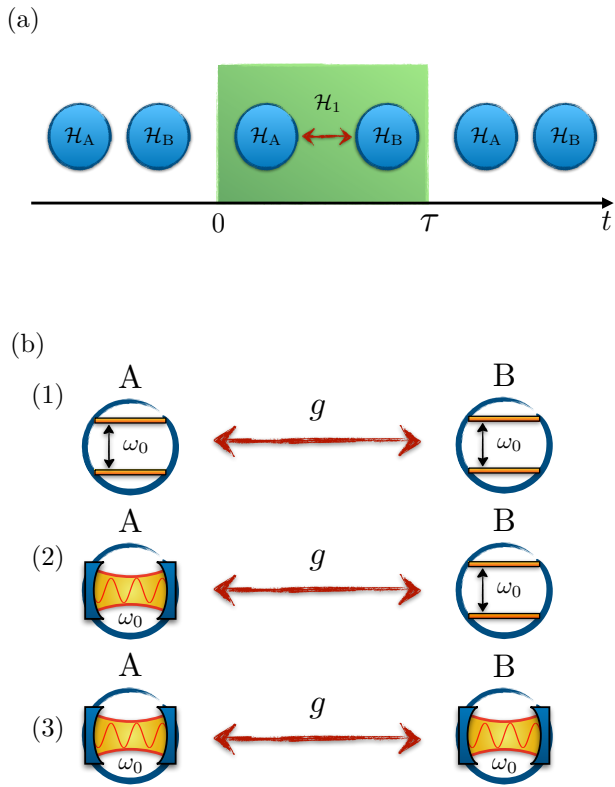

In this context, a number of researchers has been working on ``quantum batteries''~\cite{Alicki13, Hovhannisyan13, Binder15, Campaioli17,Ferraro17,Le17,Serra18}, i.e.~quantum mechanical systems for storing energy where genuine quantum effects can be used to obtain more efficient and faster charging processes with respect to classical analog systems. From an abstract point of view, the fact that quantum coherent processes can be faster then classical operations is a known fact emerging from quantum information theory and, specifically, from the concept of quantum speed limits~\cite{Giovannetti2003a,Giovannetti2003b,Giovannetti2004}. The idea of exploiting quantum coherence for efficiently charging (or discharging) quantum batteries has been studied in a fully abstract fashion~\cite{Alicki13, Hovhannisyan13, Binder15, Campaioli17}, and, more recently, by expoiting concrete models that can be implemented in laboratories~\cite{Ferraro17,Le17}. 
 
In this Article we follow the same research line but, differently from  previous attempts~\cite{Ferraro17,Le17}, we focus only on minimal models of quantum batteries, which can be solved exactly. The simplicity of our toy models allows us, on the one hand,  to avoid all subtle approximations and formal technicalities needed to handle more sophisticated models such as those studied in Refs.~\onlinecite{Ferraro17,Le17}, and, on the other hand, to identify general features, which are independent of the details of the specific experimental implementation.

To this end, we model a quantum battery as either a two-level system (TLS) or a quantum harmonic oscillator (QHO), the same simplified picture being also used for the charging system---see Fig.~\ref{fig:System}a) and b). The basic idea here is that TLSs and QHOs can be viewed as elementary building blocks of more complex quantum batteries. Also, the models considered in this work can be experimentally implemented. Indeed, TLSs and QHOs are ubiquitous in atomic and condensed matter physics. They are elementary building blocks of cavity QED architectures~\cite{Haroche_Book,Haroche13} and in systems of trapped ions~\cite{Blatt08,Kielpinski02}, ultracold atoms~\cite{Ketterle02,Lewenstein07}, superconducting circuits~\cite{Devoret13,Schoelkopf08,blais_pra_2004,Clarke08}, and semiconductor quantum dots~\cite{Tapash1999,Stockklauser17,singha_science_2011,polini_naturenano_2013,Kung12,Rossler15,hensgens_arxiv_2017,Samkharadze18}.

For the three charger-battery combinations illustrated in Figs.~\ref{fig:System}b) and by means of a unitary Hamiltonian interaction, we study energy transfer processes from the charger initialized in an arbitrary state to the quantum battery initialized in the ground state. We are particularly interested in understanding the relevance of quantum coherence for improving the efficiency of the charging process and in clarifying the role of coupling terms that do not commute with the local Hamiltonians of the model. 
  Among the main results of this Article, we emphasize the following ones: i) when a TLS-based quantum battery is charged via a QHO it is convenient to prepare the charger in a Fock state which, for sufficiently large energies, can be safely replaced by a coherent state giving approximately equal performances; ii) in the previous situation, we observe that the charging time is inversely proportional to the square root of the charger energy.

In our treatment, we focus on average energies (i.e.~Hamiltonian expectation values) without taking into account statistical fluctuations. For this reason, the presented approach is applicable also in contexts other than that of quantum batteries, such as that of heat transport processes \cite{Kurizki13}. 

Our Article is organized as follows. A general theory of energy transfer and different  models of quantum batteries are presented in Sect.~\ref{Theory of energy transfer} for the special case where the coupling Hamiltonian between the charger and the battery preserve the local energy of the system. This analysis  is then extended to non-commuting Hamiltonians (i.e.~going beyond energy-preserving protocols) in Sect.~\ref{Theory of energy transfer in  the non commuting case}. 
 A brief summary and our main conclusions are finally reported in Sect.~\ref{Conclusions}. Useful technical details can be found in Appendix~\ref{appendixRmatrix}.

\section{Energy transfer in the charger-quantum battery setup}
\label{Theory of energy transfer}

In this Section we introduce a general theoretical framework to address the charging process of a quantum battery schematically represented in Fig.~\ref{fig:System}.
We consider two quantum systems, $\rm A$ and $\rm B$, where $\rm A$ is the ``charger'',  initially containing some input energy, while $\rm B$ is the proper ``quantum battery'', initially prepared in the ground state. We denote by $\rho_{\rm A}(t)$ and $\rho_{\rm B}(t)$ the density matrices representing  their respective quantum states and with $\mathcal{H}_{\rm A}$ and $\mathcal{H}_{\rm B}$ the corresponding time-independent local Hamiltonians. We can therefore identify with $E_{\rm A}(t)={\rm tr} [ \mathcal{H}_{\rm A} \rho_{\rm A}(t)]$ the energy of the charger and with $E_{\rm B}(t)={\rm tr} [ \mathcal{H}_{\rm B} \rho_{\rm B}(t)]$ the energy of the quantum battery.
We assume that at time $t=0$ the charger is initialized in an arbitrary state while the battery is in its ground state, i.e.
\begin{equation}\label{initial state}
\rho_{\rm AB}(0)=\rho_{\rm A}(0) \otimes  | 0\rangle \langle 0|_{\rm B}~,
\end{equation}
such that $E_{\rm A}(0)>0$ and $E_{\rm B}(0)=0$.
We model the charging process as the physical operation of letting A and B interact for a finite amount of time $\tau$, as in Fig.~\ref{fig:System}a). More precisely, we assume the following global Hamiltonian
\begin{equation}\label{protocol}
\mathcal{H}(t)=\mathcal{H}_0+\lambda(t)\mathcal{H}_1~,
\end{equation}
where $\mathcal{H}_0=\mathcal{H}_{\rm A}+\mathcal{H}_{\rm B}$, $\mathcal{H}_1$ is some given interaction Hamiltonian, and $\lambda(t)$ is a dimensionless coupling constant, equal to $1$ for $t\in[0,\tau]$ and $0$ elsewhere. Physically, the ``on/off" coupling constant $\lambda(t)$ is the only classical parameter which can be externally controlled. (This can be implemented using a quantum clock, see e.g.~Ref. ~\onlinecite{Ito17}). This implies that the total energy $E(t)={\rm tr} [ \mathcal{H}(t) \rho_{\rm AB}(t)]$ is constant at all times with the exception of the switching times, i.e.~$t=0$ and $t=\tau$, where some non-zero energy can be exchanged, representing the thermodynamic work cost of switching on and off the interaction. Such cost can be quantified as the total energy change  at both switching points, i.e.
\begin{align}\label{protocol-cost}
\delta E_{\rm sw}(\tau) &\equiv [E(\tau_+)-E(\tau_-)] + [E(0_+)-E(0_-)]\nonumber \\
          &= {\rm tr}\left\{ {\cal H}_1 \left[\rho_{\rm AB}(0)-\rho_{\rm AB}(\tau) \right]\right\}~, 
\end{align}
where 
$\rho_{\rm AB}(\tau)= e^{-i({\cal H}_0 + {\cal H}_1)\tau}
 \rho_{\rm AB}(0)e^{i({\cal H}_0 + {\cal H}_1)\tau}$  ($\hbar = 1$ throughout this Article).

We first consider the case in which the interaction Hamiltonian commutes with the sum of the local terms,
\begin{eqnarray}\label{commutativity}
[\mathcal{H}_{0},\mathcal{H}_{1}]=0~,
\end{eqnarray}
ensuring $\delta E_{\rm sw}(\tau)=0$ for every initial state. From a physical point of view, this choice corresponds to energy-preserving protocols in which all the energy stored in the quantum battery B at the end of the charging process originates, without any thermodynamic ambiguity, from the charger A. In this case, the performances of the charger-battery setup can be studied in terms of the 
(mean) energy stored in the battery and the corresponding average storing power, defined respectively as
\begin{eqnarray}\label{stored energy}
E_{\rm s}(\tau)&\equiv& E_{\rm B}(\tau)={\rm tr}[\mathcal{H}_{\rm B} \rho_{\rm B}(\tau)]\;, \\ \label{storing power}
P_{\rm s}(\tau)&\equiv&  E_{\rm s}(\tau)/\tau~.
\end{eqnarray}
Upon optimization with respect to the charging time $\tau$, we can extract from these functionals a collection of {\it figures of merit} which quantify the ``quality'' of a given charging protocol from different perspectives. Specifically, we define 
the maximum (mean) energy that can be stored in the quantum battery 
\begin{equation}\label{max energy}
\overline{E}_{\rm s}\equiv \max_\tau [E_{\rm s}(\tau)]\equiv E(\overline \tau)~,
\end{equation}
the maximum power,
\begin{equation}\label{max storing power}
\tilde P_{\rm s}\equiv \max_\tau [P_{\rm s}(\tau)]~,
\end{equation}
and their corresponding  optimal charging times
\begin{eqnarray}\label{storing time} 
\overline{\tau}\equiv \min_{E(\overline \tau)= \overline{E}_{\rm s}} (\tau)\;, \qquad 
\tilde{\tau}\equiv \min_{P(\tilde \tau)= \tilde{P}_{\rm s}} (\tau)~.
\end{eqnarray}
Finally, we also introduce the  charging power at maximum energy,
\begin{equation}\label{storing power at max energy}
\overline{P}_{\rm s} \equiv \overline{E}_{\rm s}/\overline{\tau}= E_{\rm s}(\overline{\tau})/\overline{\tau}~,
\end{equation}
which, due to the fact that $\overline{\tau}$ and $\tilde{\tau}$ may not necessarily coincide, will in general
be smaller than $\tilde{P}_{\rm s}$.
 
For non-commuting interactions $[\mathcal{H}_{0},\mathcal{H}_{1}]\neq0$ more caution should be used when defining the figures of merit for a given charging protocol. Indeed, in this case, the final energy of the quantum battery will not come only from the charger A but also from the classical modulation of the coupling constant $\lambda(t)$ and, for this reason, the ``quality'' of the protocol has some degree of arbitrariness depending on which of the two energy fluxes is actually desired. The analysis of this particular situation is postposed to Sect.~\ref{Theory of energy transfer in  the non commuting case}. In the  next  Section, instead, we study
$E_{\rm s}(\tau)$ and $P_{\rm s}(\tau)$ 
for three alternative models of the charger-battery
setting that fulfill the commutativity identity~(\ref{commutativity}) and admit full analytical treatment, looking for the presence of advantages associated with the quantum structure of the system dynamics.
As a useful tool for this analysis, we compare the optimal charging times~(\ref{storing time}) to the 
quantum speed limit (QSL) time $\tau_{\rm QSL}$~\cite{Giovannetti2003a,Giovannetti2003b,Giovannetti2004,Defner17} that defines the  minimum temporal interval needed to let a quantum system to evolve between two orthogonal states under the action of its (time-independent) Hamiltonian $\mathcal{H}$, i.e.
\begin{equation}\label{eq:speedlimit}
\tau_{\rm QSL }= \frac{ \pi}{2} \frac{1}{{\rm min} \{\braket{\mathcal{H}},\braket{\delta\mathcal{H}}  \}  }~,
\end{equation}
with  $\braket{\mathcal{H}}$  indicating the gap between 
the mean value and the ground-state energy of $\mathcal{H}$, evaluated on the system input state, and  $\braket{\delta\mathcal{H}}$ being instead 
the corresponding square root of the variance of $\mathcal{H}$.

\subsection{Energy transfer between two TLSs}\label{TLS-TLS}

\begin{figure}[t]
\centering
\begin{overpic}[width=1\columnwidth]{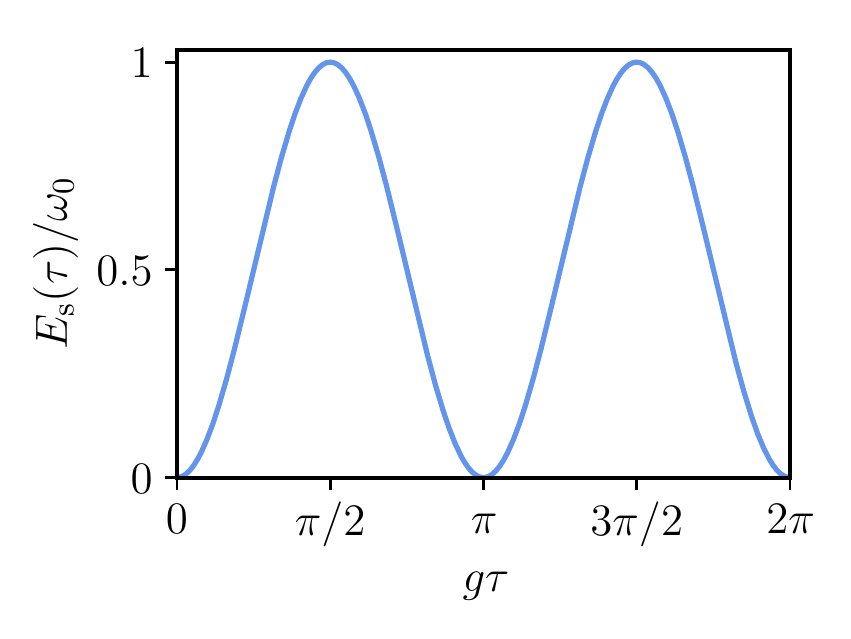}\put(3,66){\normalsize (a)}\end{overpic}\vspace{0.5em}
\begin{overpic}[width=1\columnwidth]{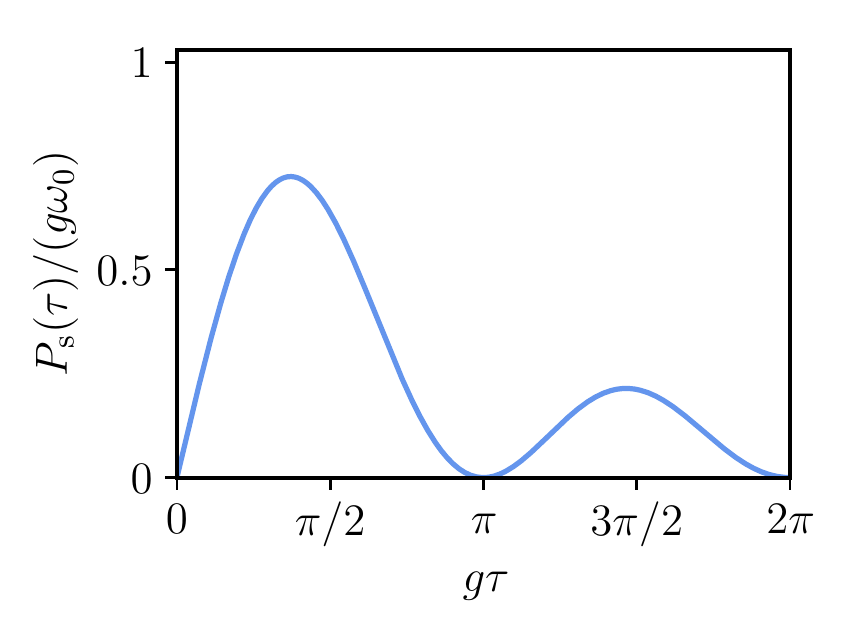}\put(3,66){\normalsize (b)}\end{overpic}
\caption{(Color online) Panel (a) displays the stored energy $E_{\rm s}(\tau)$ (in units of $\omega_{0}$) as a function of $g\tau$, for the case of two coupled qubits. Panel (b) shows the average charging power $P_{\rm s}(\tau)$ (in units of $g\omega_{0}$) as a function of $g \tau $. We clearly see that the quantum battery is charged by the charging qubit via Rabi oscillations.\label{fig:Eqq}}
\end{figure}

We begin by studying the simplest, yet non-trivial, case of a charger-battery setting which we will use as reference 
for the following study. Here, the  charger and  quantum battery are two resonant TLSs (also named qubits throughout this Article), coupled via an energy-preserving interaction that merely shifts excitation quanta between the two qubits. Accordingly, we write the system Hamiltonian~(\ref{protocol}) in terms of the following components:
\begin{eqnarray}\label{H_qq}
\mathcal{H}_{\rm A}&=&\frac{\omega_0}{2}  \left(\sigma_z^{(\rm{A})}+1\right)~, \\  \mathcal{H}_{\rm B}&=&\frac{\omega_0}{2} \left(\sigma_z^{(\rm{B})}+1\right)~, \nonumber \\ \mathcal{H}_1&=&g\left(\sigma_-^{(\rm{A})}\sigma_+^{(\rm{B})}+\sigma_+^{(\rm{A})}\sigma_-^{(\rm{B})}\right) ~, \nonumber
\end{eqnarray}
where $\omega_0$ is the level spacing of each TLS, $\sigma_z^{(\rm{S})}$  are Pauli matrices acting on the S = A,B subspaces, $\sigma_+^{(\rm{S})},\sigma_-^{(\rm{S})}$ are spin ladder operators acting on the same subspaces, and $g$ is the coupling strength.
 In this case, energy transfer is occurring through the well-known Rabi oscillations, see Fig.~\ref{fig:Eqq}. Indeed, exploiting the fact that Eq.~(\ref{commutativity}) holds, one can easily show that, assuming the charger A to be initialized in the excited state $\ket{1}_{\rm A}$ and the qubit B in the ground state $\ket{0}_{\rm B}$, the evolved system can be expressed as
\begin{eqnarray}\label{stateqq}
\ket{\Psi(t)}_{\rm AB}=& e^{-i\omega_0t}&\big[\cos(gt)\ket{1}_{\rm A}\ket{0}_{\rm B}\\ &-&i\sin(gt)\ket{0}_{\rm A}\ket{1}_{\rm B} \big] ~,\nonumber
\end{eqnarray}
yielding 
\begin{eqnarray}\label{stored energyqq}
E_{\rm s}(\tau)=\omega_0 \sin^2(g\tau)~, \quad 
P_{\rm s}(\tau)=\omega_0 \frac{\sin^2(g\tau)}{\tau}~,
 \end{eqnarray}
 for the quantities (\ref{stored energy}) and  (\ref{storing power}).
The maximum energy is hence provided by $\overline{E}_{\rm s}=\omega_0$ and is achieved at time $\bar{\tau}=\pi/(2g)$ (the corresponding
power at maximum energy transfer (\ref{storing power at max energy}) being $\overline{P}_{\rm s}=2 g \omega_0/\pi$). 
 The maximum power instead is  $\tilde{P}_{\rm s}\approx 0.72 g\omega_0$  and is achieved at time $\tilde{\tau}\approx 1.16/g$ (result obtained 
by simple numerical inspection of the function $y=\sin^2(x)/x$, which has maximum value $\tilde{y}\approx 0.72$ at $\tilde{x}\approx1.16$).

\subsection{Energy transfer between a QHO and a TLS battery}\label{TLS-QHO}
\begin{figure*}[t]
	\centering
	\begin{tabular}{cc}
	\begin{overpic}[width=1\columnwidth]{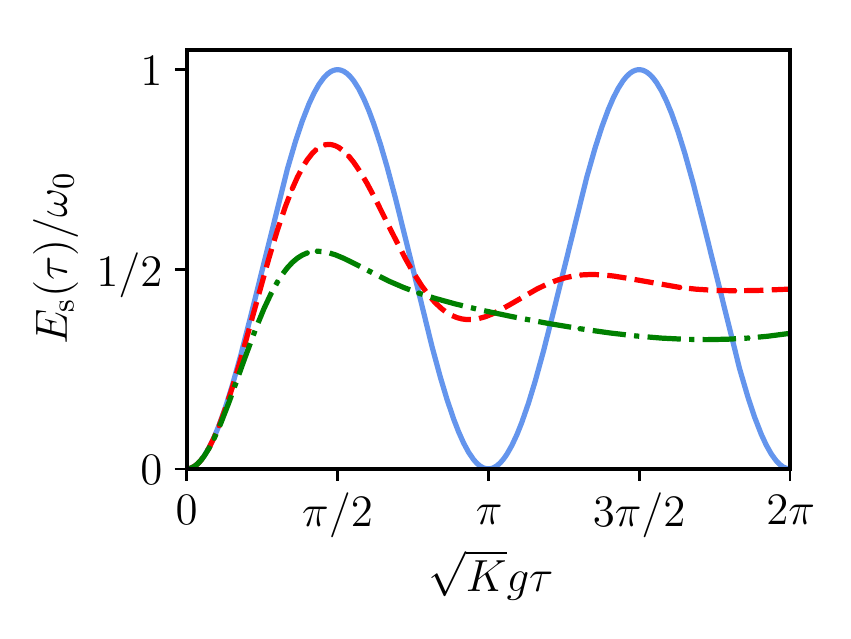}\put(3,66){\normalsize (a)}\end{overpic}\vspace{0.5em} & 
	\begin{overpic}[width=1\columnwidth]{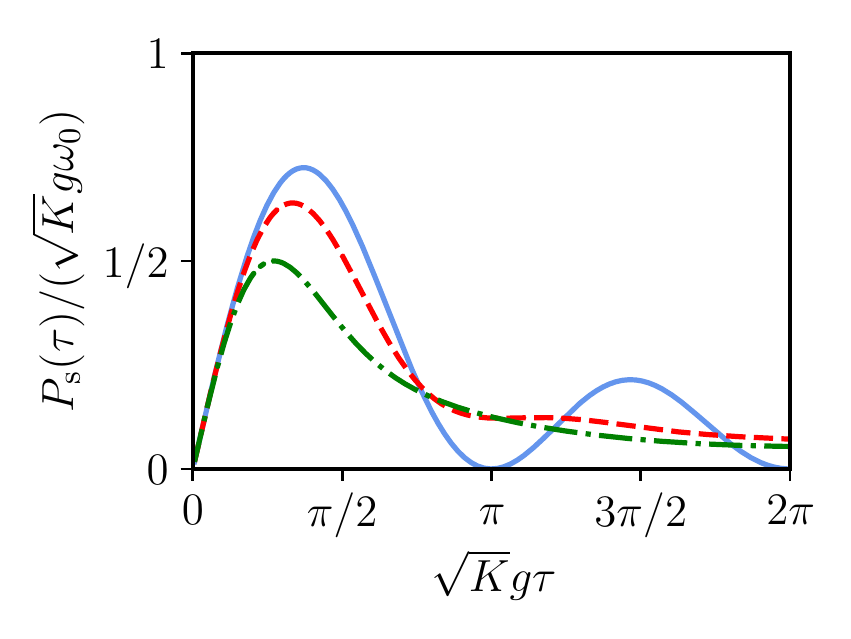}\put(3,66){\normalsize (b)}\end{overpic}\vspace{0.5em} \\
\begin{overpic}[width=1\columnwidth]{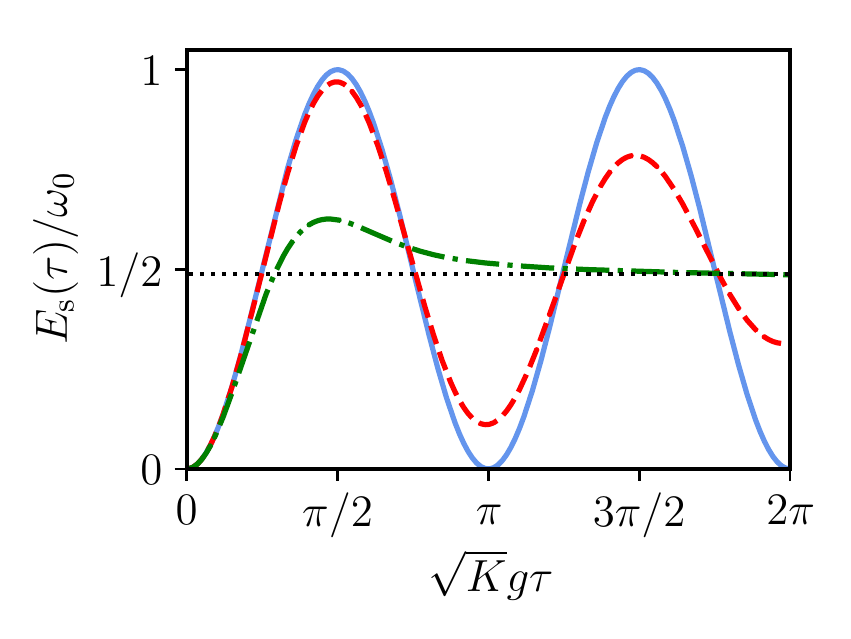}\put(3,66){\normalsize (c)}\end{overpic}\vspace{0.5em} & 
\begin{overpic}[width=1\columnwidth]{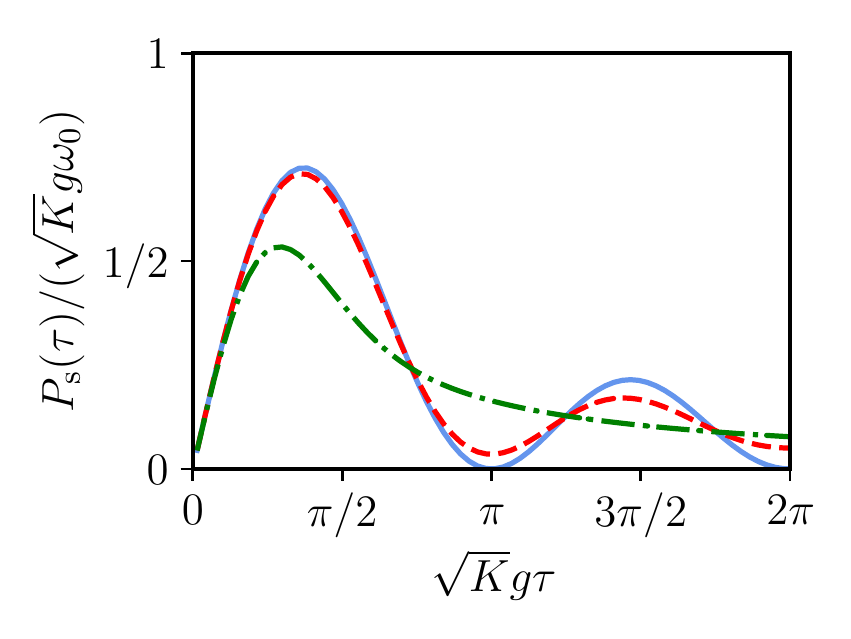}\put(3,66){\normalsize (d)}\end{overpic}\vspace{0.5em}
\end{tabular}
\caption{(Color online)  Panel (a) displays the stored energy $E_{\rm s}(\tau)$ (in units of $\omega_{0}$) as a function of $\sqrt{K}g \tau$, for the case of a qubit charged by a QHO. The initial number of excitation is $K=3$. Different curves refer to results obtained for three different choices of the initial state of the charger: Fock state (blue solid line), coherent state (red dashed line), and Gibbs state (dash-dotted green line). Panel (b) shows the average charging power $P_{\rm s}(\tau)$ (in units of $\sqrt{K} g \omega_{0}$) as a function of $\sqrt{K}g \tau$. The initial number of excitation is $K=3$. Color coding as in panel (a).  Panel (c) same as in panel (a) but for $K=20$. The black dotted line represents the energy of a Gibbs state of the qubit, with temperature equal to that of the initial Gibbs state of the QHO. Note that, for long times (not shown), the system has revivals, due the unitarity of the time evolution. Panel (d) shows the average charging power corresponding to panel (c). This figure clearly shows the ``optimality'' of the Fock state, which is the best choice for maximizing the energy and power. Note that, for $K\gg1$ as in panels (c) and (d), the coherent state well approximates the quantities $E_{\rm s}(\tau)$ and $P_{\rm s}(\tau)$ calculated for an initial Fock state.\label{fig:Eqo}}
\end{figure*}

We now focus on the case in which the charger A is described by a QHO while the quantum battery B is still described by a TLS. The relevant Hamiltonians are
\begin{eqnarray}\label{H_qo}
\mathcal{H}_{\rm A}&=&\omega_0 a^\dagger a~, \\ \mathcal{H}_{\rm B}&=&\frac{\omega_0}{2} \left(\sigma^{(B)}_z+1\right)~,  \nonumber\\ \mathcal{H}_1&=&g\left( a^\dagger \sigma^{(B)}_-+ a\sigma^{(B)}_+\right) ~, \nonumber
\end{eqnarray}
where $a^\dagger$ ($a$) is the creation (destruction) bosonic operator acting on A,  and where again
$\omega_0$ and $g$ are respectively the characteristic frequency of both systems and 
the coupling strength parameter of the model. 
The model described by the total Hamiltonian  $\mathcal{H}=\mathcal{H}_{\rm A}+\mathcal{H}_{\rm B}+\mathcal{H}_{1}$ is the so-called Jaynes-Cumming model~\cite{Jaynes63}, which can be seen as the building block of much more complicated many-body models such as the Tavis-Cummings~\cite{Tavis68,Tavis69} and Dicke models~\cite{Dicke54}.

We now note that from the commutativity relation~(\ref{commutativity}) the operator $K=a^\dagger a+\sigma_z/2$, which counts the total number of excitations, commutes with the full Hamiltonian $\mathcal{H}$ and is therefore a constant of the motion. We can  hence solve the dynamics by restricting the analysis to subspaces with a given number $n$ of excitations spanned by the vectors $\ket{n}_{\rm A}\ket{0}_{\rm B}$ and $\ket{n-1}_{\rm A}\ket{1}_{\rm B}$, 
where Hamiltonian simplifies to the one described in the previous Section---see Eq.~(\ref{H_qq})---with appropriate renormalized parameters. 
Here the eigenvectors of ${\cal H}$ are $\ket{\pm,n}=(\ket{n}_{\rm A}\ket{0}_{\rm B}\pm\ket{n-1}_{\rm A}\ket{1}_{\rm B})/\sqrt{2}$ and the corresponding eigenvalues are $\omega_{\pm,n}=n \omega_0\pm \sqrt{n}g$.  
Therefore if we start from the initial quantum state $\ket{n}_{\rm A}\ket{0}_{\rm B}$, its temporal evolution is given by
\begin{eqnarray}
 |\Psi_n(t) \rangle_{\rm AB}=& e^{-in\omega_0 t} & \big[\cos(\sqrt{n}gt)\ket{n}_{\rm A}\ket{0}_{\rm B} \nonumber\\ & - & i\sin(\sqrt{n}gt) \ket{n-1}_{\rm A}\ket{1}_{\rm B}\big]~.\label{DYNA} 
\end{eqnarray}
Consider next the
case of a generic input of the form ~(\ref{initial state}) where we fix the initial energy  to the value $E_{A}(0)=E_{\rm in}$ and hence 
 the average number of excitations to $K=E_{\rm in}/\omega_0$.
Expanding $\rho_{\rm AB}(0)$ on the Fock basis $\ket{n}_{\rm A}\ket{0}_{\rm B}$, from Eq.~(\ref{DYNA}) 
we can calculate the mean stored energy and the average charging power:
\begin{equation} \label{EQ1new} 
E_{\rm s}(\tau)=\omega_0 \sum_n p_n^{(K)} \sin^2(\sqrt{n}g\tau)\;,
\end{equation}
and
\begin{equation}\label{EQ2new} 
P_{\rm s}(\tau)=\frac{\omega_0 \sum_n p_n^{(K)} \sin^2(\sqrt{n}g\tau)}{\tau}~,
\end{equation}
where $p_n^{(K)}$ is the diagonal part of $\rho_{\rm A}(0)$ in the Fock basis, subject to the constraint of yielding the selected initial energy, i.e.~$\sum_n  n p_n^{(K)}=K$.  

Let us first study the case of an initial state of the Fock type. In this case, $p_K^{(K)}=1$ and $p_{n \neq K}^{(K)}=0$, and Eqs.~(\ref{EQ1new}) and~(\ref{EQ2new}) become
\begin{eqnarray}\label{storedEP}
E^{\rm F}_{\rm s}(\tau)&=&\omega_0 \sin^2(\sqrt{K}g\tau)\;, \\ 
 \label{STOREPF} 
P^{\rm F}_{\rm s}(\tau)&=&\omega_0 \frac{\sin^2(\sqrt{K}g\tau)}{\tau}~,
\end{eqnarray}
where ``F'' denotes that the initial state of the charger is a Fock state. The maximum of Eq.~(\ref{storedEP}) is $\overline{E}^{\rm F}_{\rm s}=\omega_0$ and is achieved for the first time at
\begin{eqnarray} \label{eq:quantum_advantage}
\bar{\tau}=\pi/(2\sqrt{K}g)~.
\end{eqnarray}
At this special time the battery gets completely charged, resulting in a final 
state of the AB system that  exactly factorizes, i.e. $|K-1\rangle_{\rm A} |1\rangle_{\rm B}$. Due to the properties of the function $\sin^2{(x)}/x$---Sect.~\ref{TLS-TLS}---the maximum value of the  power  (\ref{STOREPF}) is instead provided by $\tilde{P}^{\rm F}_{\rm s}\approx 0.72 g\omega_{0}\sqrt{K}$  and is achieved at time
\begin{eqnarray}\label{eq:q_adv1} 
\tilde{\tau}\approx 1.16/(\sqrt{K} g)~,
\end{eqnarray}
which, apart from a multiplicative constant, exhibits the same $1/\sqrt{K}$  scaling of Eq.~(\ref{eq:quantum_advantage}). Compared with the two qubits model of the previous Section, Eq.~(\ref{stored energyqq}), in the present case there is still a transfer of only one quantum of energy  from A to B but in a time window that is reduced by a factor $1/\sqrt{K}$. Thus we can say that, from the initial number $K$ of excitations in the system, only one is eventually transferred from the charger to the quantum battery, with the other $K-1$ ones acting as a catalytic resource that increases the speed of the process.
 This cooperative effect is the one that ultimately leads to the $1/\sqrt{K}$ improvement reported in Eq.~(\ref{eq:quantum_advantage}) which,
despite the lack of collective behavior stemming from the mutual interactions between $K$ qubit batteries coupled to  
a single common photonic mode, mimics  a similar scaling observed  in Ref.~\onlinecite{Ferraro17}.
 Such advantage can also be connected with the QSL bound~(\ref{eq:speedlimit}) confirming an argument of Ref.~\onlinecite{Campaioli17}. Indeed, by direct evaluation, we have $\braket{\mathcal{H}}=K\omega_0$ and $\braket{\delta\mathcal{H}}=\sqrt{\braket{\delta\mathcal{H}_{\rm A}}^2+Kg^2}\geq \sqrt{K}g$, which, for  $K$ big enough, gives $\tau_{\rm QSL}\simeq  {\pi}/({2  \sqrt{K}g})$ 
reproducing the scaling of Eq.~(\ref{eq:quantum_advantage}).

Consider next the case where A is initialized in a generic (not necessarily Fock) input state. From a close inspection of Eqs.~(\ref{EQ1new})
and (\ref{EQ2new}) 
it turns out that, for fixed $K$, 
 the values of $\overline{E}^{\rm F}_{\rm s}$ and  $\tilde{P}^{\rm F}_{\rm s}$ are bigger than the corresponding quantities one can obtain with any other input state of A having  the same expectation value of the input energy of the selected Fock state. Indeed, from Eq.~(\ref{EQ1new}) we have 
$E_{\rm s}(\tau)\leq\omega_{0}=\overline{E}^{\rm F}_{\rm s}$, while from Eq.~(\ref{EQ2new}) we obtain
\begin{eqnarray}
P_{\rm s}(\tau)&=& \omega_0 g \; \sum_n  \sqrt{n} ~p_n^{(K)} \left[ \frac{ \sin^2(g\sqrt{n}\tau)}{g \sqrt{n}\tau} \right] \nonumber\\
&\leq& \omega_0 g \max_x \left[\frac{ \sin^2(x)}{x} \right] \sum_n  \sqrt{n} ~ p_n^{(K)}  \nonumber  \\
\label{eq:final_step}
 &\leq& \omega_{0} g \sqrt{K}  \max_x \left[\frac{ \sin^2(x)}{x} \right] =\tilde{P}^{\rm F}_{\rm s}~,
\end{eqnarray}
where in the second inequality we used the concavity of the function $\sqrt{x}$ to write  $\sum_n p_n^{(K)}\sqrt{n}\leq \sqrt{K}$. 
These relations are also evident in Fig.~\ref{fig:Eqo} where we plot 
the stored energy $E^{\rm F}_{\rm s}(\tau)$ and the average charging power $P^{\rm F}_{\rm s}(\tau)$ of the Fock input case, together with the corresponding 
values of 
$E_{\rm s}(\tau)$ and $P_{\rm s}(\tau)$ obtained for different choices of the input state of A 
(namely the case of a coherent input and the one of a thermal distribution, characterized by a Poissonian distribution $p_n=e^{-K}K/n!$ and a Gibbs distribution $p_n=[K/(K+1)]^n/(K+1)$, respectively).

According to the above analysis,
 for fixed input mean energy of the charger A, Fock states provide optimal performances with respect to all our figures of merit. A Fock state, however, is not always easy to be prepared experimentally~\cite{Hofheinz08} for an arbitrary number of photons $K$. One may therefore be interested in replacing it with a more affordable coherent state $|\sqrt{K}\rangle$ having the same energy. Luckily, from our previous formulas (see also Fig.~\ref{fig:Eqo}) it is evident that for $K\gg 1$, a coherent state and a Fock state produce almost indistinguishable results. More generally, this fact is valid for every initial state with a sufficiently peaked energy distribution $\{ p_n^{(K)}\}_n$, i.e.~a state such that $\langle (a^\dag a)^2 \rangle \ll \langle a^\dag a \rangle$. Such weak dependence on the specific initial state is clearly crucial for the purpose of validating experimentally the $1/\sqrt{K}$ scaling
of the optimal charging times reported in  Eqs.~(\ref{eq:quantum_advantage}) and (\ref{eq:q_adv1}). 

Finally, we note that the role of quantum coherence is not crucial in the charging step of a quantum battery. Indeed, Fock states, which provide optimal performances, have no coherence in the basis of the eigenstates of the Hamiltonian of the charger. Furthermore, because of Eq.~(\ref{eq:final_step}), any coherent combination of Fock states is not optimal.  The role of entanglement is more much subtle and is thoroughly discussed in Ref.~\onlinecite{andolina_arxiv_2018}.

\subsection{Energy transfer between two QHOs}\label{subsection_oo}

We now study the case in which both A and B are QHOs with a quadratic Hamiltonian ${\cal H}$ characterized by  the following terms:
\begin{eqnarray}\label{H_oo}
\mathcal{H}_{\rm A}&=&\omega_0 a^\dagger a~, \\
\mathcal{H}_{\rm B}&=&\omega_0 b^\dagger b~, \nonumber \\ 
\mathcal{H}_1&=&g(a^\dagger b+a b^\dagger)~. \nonumber
\end{eqnarray} 
The operator $\mathcal{H}_{\rm A}+ \mathcal{H}_{\rm B} + \mathcal{H}_{1}$ can be diagonalized in terms of the ``normal'' bosonic operators, $\gamma_{\pm}=(a\pm b)/\sqrt{2}$, with associated normal frequencies $\omega_{\pm}=\omega_0\pm g$ which, to guarantee overall stability, are taken positive by assuming $|g|\leq \omega_0$.

As usual, we fix the initial mean energy  of the charger A ($E_{A}(0)=E_{\rm in}$)  and define the average number of excitations, $K=E_{\rm in}/\omega_0$. 
 In order to calculate the stored energy~(\ref{stored energy}) we find then useful to adopt  the Heisenberg representation writing $E_{\rm s}(\tau) ={\rm tr}[\rho_{\rm AB}(0)\mathcal{H}_{\rm B}(\tau)]$, with $\mathcal{H}_{\rm B}(\tau)\equiv e^{i \mathcal{H} \tau} \mathcal{H}_{\rm B}e^{-i \mathcal{H} \tau}$. Expressing hence $a$ and $b$ as functions of the normal operators $\gamma_{\pm}$ and using that the latter evolve simply as $\gamma_{\pm}(t) = e^{-i\omega_{\pm}t}\gamma_{\pm}$, we obtain
 \begin{eqnarray}\label{evolvedH}
\mathcal{H}_{\rm B}(\tau)&=&\frac{\omega_0}{2}\Bigg\{a^\dagger a+b^\dagger b \\&-&\left[\frac{e^{-i2g\tau}}{2} (a^\dagger a-b^\dagger b+b^\dagger a-a^\dagger b) +{\rm H.c.}\right]    \Bigg\}~. \nonumber
 \end{eqnarray} 
This considerably simplifies the calculation of $E_{\rm s}(\tau)$ since the initial state contains no excitations on B, 
yielding 
\begin{eqnarray}\label{stored energyoo}
E_{\rm s}(\tau)&=&K\omega_0 \sin^2(g\tau)~, \\ \label{powerstored1} 
P_{\rm s}(\tau)&=&\frac{K\omega_0 \sin^2(g\tau)}{\tau}~,
 \end{eqnarray} 
the formulas applying irrespectively from the details of the initial state 
 (a direct consequence of the quadratic form of the Hamiltonian, for which the dynamics of the first and second moments---e.g.~$\braket{a} , \braket{b}, \braket{a^\dagger a} $, etc---is independent of higher-order ones). 

Equations~(\ref{stored energyoo}) and~(\ref{powerstored1}) have exactly the same dependence on time of Eq.~(\ref{stored energyqq}) for the case of two TLSs model. Hence, the optimal charging times of the two models coincide, i.e.~$\bar{\tau}=\pi/(2g)$ and $\tilde{\tau}\approx 1.16/g$, and exhibit no speedup in $K$. Nonetheless, 
due to the higher storing capability of the QHO battery which has now an unbounded energy spectrum, in the present case the values for the associated maximal stored energy and
maximal power (i.e.~$\overline{E}_{\rm s}=K\omega_0$ 
and  $\tilde{P}_{\rm s}\approx 0.72 gK\omega_0$) show a linear increase in $K$ that was absent in the model of Sect.~\ref{TLS-TLS}.
It is also worth stressing that the $K$ improvement for $\tilde{P}_{\rm s}$ reported here has a completely different origin
with respect to the $\sqrt{K}$ power improvement observed  in Sect.~\ref{TLS-QHO}. Indeed, due to the absence of an unbounded energy spectrum for the battery of the QHO-TLS model, the $\sqrt{K}$ improvement of the previous Section is just a consequence of the speedup in the charging time (\ref{eq:q_adv1}) which, as already noticed,  
is instead absent in the present model.  
The value of $\bar{\tau}=\pi/(2g)$ obtained here, can finally be compared with the QSL time of Eq.~(\ref{eq:speedlimit}). An analogous calculation of Sect.~\ref{TLS-QHO} gives $\tau_{\rm QSL }\simeq  \pi/(2\sqrt{K}g)$ in the large $K$ limit, 
revealing that, at variance with the QHO-TLS case, the observed $\bar{\tau}$ does not saturate the QSL bound.
This is due to the fact that, before reaching a state of maximal charging for B, the system has to travel between a finite number of orthogonal states. While the bound can be applied for each of this transition, we should take into account that we have to travel through many orthogonal states. This simple example shows that the predictions of a quantum advantage based on a speed limit argument~\cite{Campaioli17} are not always correct independently of the specific model.

\section{Theory of energy transfer in the non-commuting case} 
\label{Theory of energy transfer in  the non commuting case}

In this Section we discuss how the process of energy exchange between a charger and a quantum battery is modified when the condition $[\mathcal{H}_{0},\mathcal{H}_{1}]=0$ is not fulfilled. In this case $\delta E_{\rm sw}(\tau)\neq 0$, meaning that the protocol described by Eq.~(\ref{protocol}) does not simply enable energy transfer from A to B, since some energy is externally injected into or extracted from the whole system, via the sudden quench of the interaction Hamiltonian. To characterize the performances of these special charger-battery models we are hence forced to introduce a new functional $E_{\rm t}(\tau)$ which, at variance with Eq.~(\ref{storing power}), accounts only for the process of energy transfer from A to B, while properly neglecting the extra energy contributions induced by the external switching of ${\cal H}_1$.

Clearly, there is a certain degree of arbitrariness in giving such definition.
 In this Article we offer the following operational definition of $E_{\rm t}(\tau)$:
\begin{itemize}
\item[1)] If $\delta E_{\rm sw}(\tau)< 0$, some energy is extracted from the system ${\rm A}+{\rm B}$, which has a ``credit'' towards the external world. We can therefore safely state that all the energy stored in B comes from A setting~$E_{\rm t}(\tau)=E_{\rm s}(\tau)$;

 \item[2)] If $\delta E_{\rm sw}(\tau)>0$, some energy is injected into the system, which has a ``debit'' towards the external world. If the energy $E_{\rm A}(\tau)$ in A is sufficient to compensate this energy debit, i.e.~if $E_{\rm A}(\tau)\geq\delta E_{\rm sw}(\tau)$, we state that the remaining energy in B is a transferred energy, $E_{\rm t}(\tau)=E_{\rm s}(\tau)$.  Otherwise, if the energy $E_{\rm A}(\tau)$ in A is not sufficient, we subtract from the energy in B the remaining amount needed to pay the debit. Therefore, the transferred energy is given by  $E_{\rm t}(\tau)=E_{\rm s}(\tau)-\left[\delta E_{\rm sw}(\tau)-E_{\rm A}(\tau)\right]$. 
\end{itemize}

Summarizing, our definition of $E_{\rm t}(\tau)$ can then be expressed as
\begin{eqnarray}\label{energytransfer}
E_{\rm t}(\tau)=E_{\rm s}(\tau)-{\rm max}\big \{0,\delta E_{\rm sw}(\tau)-E_{\rm A}(\tau) \big \}~.
\end{eqnarray}
With the help of the above quantity, 
in the remaining part of this Section we study the efficiency of the two specific cases of charger-battery models with
non commuting  $\mathcal{H}_{0}$ and $\mathcal{H}_{1}$. In the first case---Sect.~\ref{detuning}---we relax the hypothesis that the two subsystems A and B are in resonance. In this case the charging protocol does not act on the system by controlling the coupling strength $g$ between A and B. Rather, control occurs on the frequency of the subsystem A, which can be brought in resonance with B or tuned away from it. 
In the second case---Sect.~\ref{beyond}---we explicitly include into the Hamiltonian terms that do not simply transfer excitations of $\mathcal{H}_0$ between the two subsystems. These terms can be neglected when the coupling constant is small, invoking the so-called ``rotating wave approximation'' (RWA)~\cite{Schleich_Book}. Hence, this beyond-RWA regime better describes the case in which the two subsystems A and B are strongly coupled. In what follows we present a simple model having a critical point in the spectrum and we show that, near the critical point, both battery and charger are externally charged via quenches and their energy increases as a power law in time. Although strong coupling can be thought of being an obvious choice to reduce the charging time, since in this case $\tilde{\tau}\sim1/g$, below we show that this regime is not optimal in the sense that it does not fit the ideal scenario of  pure-energy-exchange between A and B.

For the sake of simplicity, both Sect.~\ref{detuning} and~\ref{beyond} deal with the case of two QHOs.

\subsection{The detuning protocol}
\label{detuning}
\begin{figure}[t]
\centering
\begin{overpic}[width=1\columnwidth]{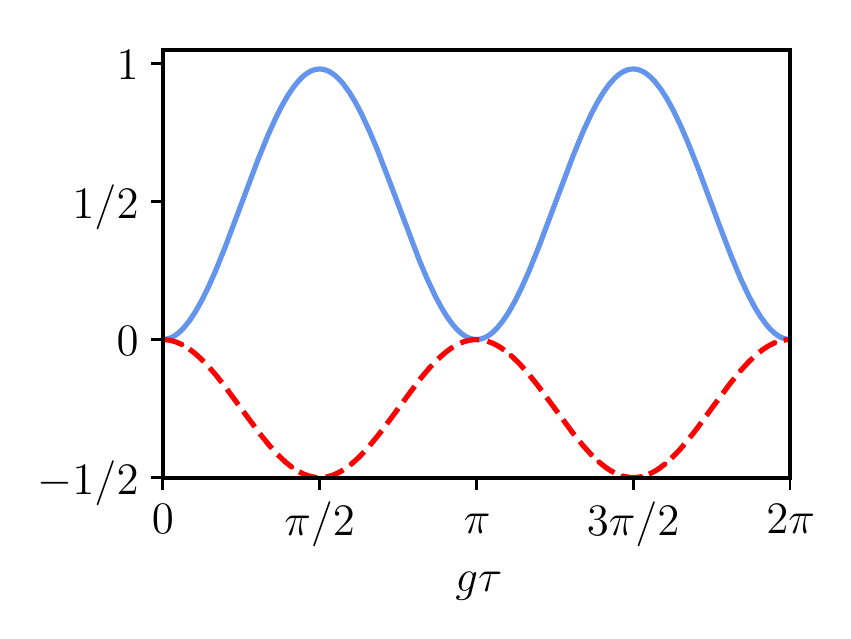}\put(3,66){\normalsize (a)}\end{overpic}\vspace{0.5em}
\begin{overpic}[width=1\columnwidth]{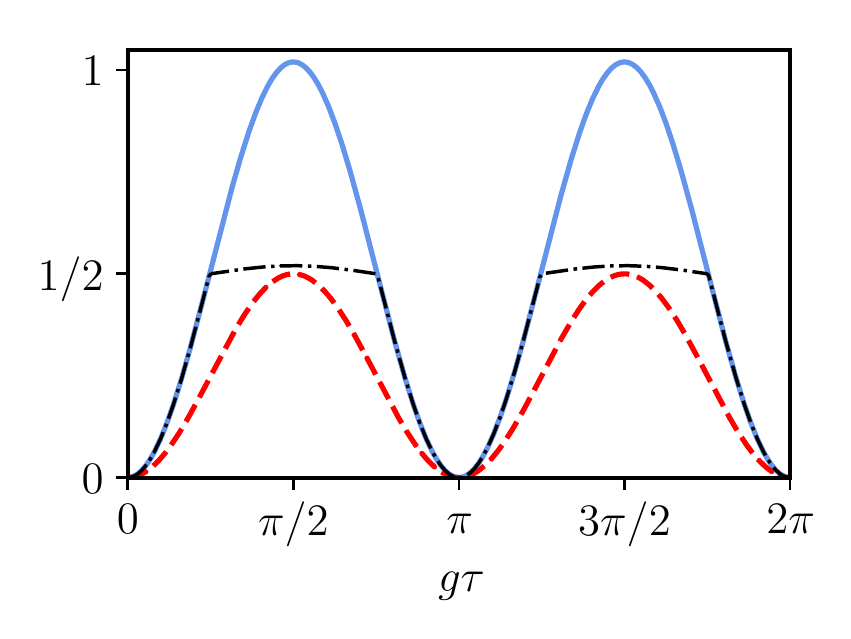}\put(3,66){\normalsize (b)}\end{overpic}
\caption{(Color online)Figures of merit for the detuning protocol described in Sect.~\ref{detuning}. Panel (a) shows the stored energy $ E_{\rm s}(\tau) $ (blue solid line) and the switching energy $\delta E_{\rm sw}(\tau)$ (red dashed line), in units of $K\omega_0$, and as functions of $g\tau$. Results in this panel have been obtained by setting $\delta\omega=\omega_0/2$ and $g=\omega_0/10$. Since $\delta\omega>0$, there is no difference between the stored energy and the transferred energy, i.e.~$E_{\rm t}(\tau)=E_{\rm s}(\tau)$. Panel (b) shows 
 the stored energy $ E_{\rm s}(\tau) $ (blue solid line),  the transferred energy $E_{\rm t}(\tau)$ (black dash-dotted line), and the switching energy $\delta E_{\rm sw}(\tau)$ (red dashed line). Results in this panel have been obtained by setting $\delta\omega=-\omega_0/2$ and $g=\omega_0/10$. Since $\delta\omega<0$, some energy is injected into the system and $E_{\rm t}(\tau)\leq E_{\rm s}(\tau)$. \label{fig:Etdetuning}}
\end{figure}

So far we have analyzed a charging protocol in which the coupling between the two subsystems A and B is turned on and off. However, this protocol may be experimentally challenging. A more practical way to control energy exchange between the two subsystems A and B consists in manipulating the frequency of the charger A, an experimentally viable route with the technology described in Ref.~\onlinecite{Hofheinz08}. The new protocol goes as following. The two subsystems A and B are initially largely detuned and energy transfer is therefore strongly suppressed. In a time window $\tau$, the detuning is set to zero and the two subsystems interact. Finally, the subsystem A is again largely detuned from B and energy flow is again blocked. Formally, the system under study consists of two QHOs with a time-dependent Hamiltonian~(\ref{protocol}) with components
\begin{eqnarray}\label{H_ooDetuning}
\mathcal{H}_{0}&=&[\omega_0+\delta \omega] a^\dagger a+\omega_0 b^\dagger b +g(a^\dagger b+a b^\dagger)~, \nonumber \\  \mathcal{H}_1&=&-\delta \omega a^\dagger a~.
\end{eqnarray} 
The quantities $\omega_{0}$ and $g$ and the operators $a$ and $b$ have the same meaning as in Eq.~(\ref{H_oo}) and $\delta\omega $ is the detuning between the two subsystems. The latter is assumed to have not a definite sign but to be large in modulus with respect to the coupling, namely  $|\delta\omega/g| \gg 1$.

We remind the reader that according to the  definition of the switching parameter 
$\lambda(t)$ of Eq.~(\ref{protocol}), 
$\mathcal{H}_{0}$ dictates the evolution at times $t^*\not\in[0,\tau]$, while $\mathcal{H}=\mathcal{H}_{0}+\mathcal{H}_{1}$ generates the evolution at time $t\in[0,\tau]$. Accordingly, at time $t^*\not\in[0,\tau]$ the two subsystems are largely detuned, and energy exchange is suppressed. Using the well-known Schrieffer-Wolff transformation~\cite{Schleich_Book}, in this time window we can effectively rewrite 
$\mathcal{H}_{0}$ as
\begin{eqnarray}\label{Heffective}
\mathcal{H}^{\rm eff}_{0}&=& \left[\omega+\delta\omega +g^2/\delta\omega \right]a^\dagger a \nonumber\\
&+&\left[\omega-g^2/\delta\omega \right]b^\dagger b
\end{eqnarray} 
up to corrections on the order of $g^3/\delta\omega^2$. This effective Hamiltonian, which is valid at all times $t^*\not\in[0,\tau]$ provided $|\delta\omega/g| \gg 1$, shows that the interaction between A and B is effectively quenched and exchange of quanta between the two subsystems is strongly suppressed. Thanks to this effective decoupling, we can define two effective local Hamiltonians acting on A and B, i.e.
\begin{eqnarray}\label{HeffectiveAB}
\mathcal{H}_{\rm A}^{\rm eff}&=&\left[\omega+\delta\omega +g^2/\delta\omega \right]a^\dagger a~,\nonumber    \\ \mathcal{H}_{\rm B}^{\rm eff}&=&\left[\omega-g^2/\delta\omega\right]b^\dagger b~,
\end{eqnarray} 
which are approximate constants of the motion. Once local Hamiltonians on A and B are defined, we can apply the general analysis described in Sect.~\ref{Theory of energy transfer} to calculate all relevant quantities. For simplicity we set $K=1$.

At times $t\in[0,\tau]$, the coupling parameter $\lambda(t)$ is equal to one and due to the presence of $\mathcal{H}_{1}$ the two subsystems are in resonance. 
In this time interval, $\mathcal{H}=\mathcal{H}_0+\mathcal{H}_1$ is identical to that reported in Eq.~(\ref{H_oo}) and, as long as we consider a density matrix of the form~(\ref{initial state}) as input state for the system,  the 
dynamical evolution can be 
described as in Sect.~\ref{subsection_oo}. Hence, it is straightforward to calculate the stored energy, the average charging power, and $\delta E_{\rm sw} (\tau)$:
\begin{eqnarray}\label{detunedpower}
E_{\rm s}(\tau)&=&\left[\omega-\frac{g^2}{\delta\omega} \right] \sin^2(g\tau) ~,\\
P_{\rm s}(\tau)&=&\left[\omega-\frac{g^2}{\delta\omega} \right] \frac{\sin^2(g\tau)}{\tau} ~, \nonumber \\
\delta E_{\rm sw} (\tau)&=&-\delta\omega\sin^2(g\tau) \nonumber ~.
\end{eqnarray} 
In the case $\delta\omega>0$ we have $\delta E_{\rm sw} (\tau)<0$ and net energy is extracted from the system. According to Eq.~(\ref{energytransfer}), $E_{\rm t}(\tau)=E_{\rm s}(\tau)$. In the case $\delta\omega >0$ net energy is injected from the outside world. In this case the transferred energy should be calculated according to the definition in Eq.~(\ref{energytransfer}) and no simplifications occur. Our main results are illustrated in Fig.~\ref{fig:Etdetuning}.

\subsection{Beyond the RWA}
\label{beyond}

\begin{figure*}[t]
	\centering
	\begin{tabular}{cc}
	\begin{overpic}[width=0.9\columnwidth]{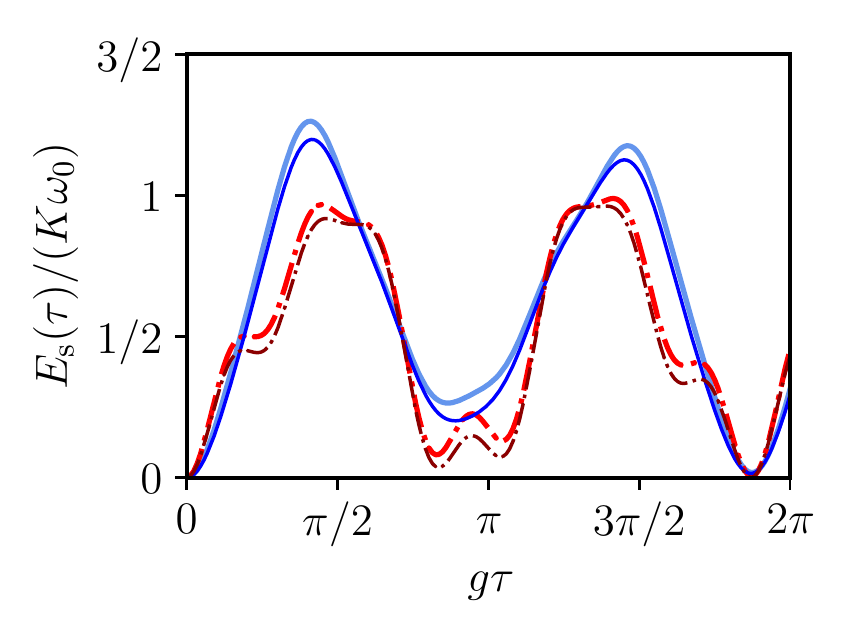}\put(3,66){\normalsize (a)}\end{overpic}\vspace{0.5em} & 
	\begin{overpic}[width=0.9\columnwidth]{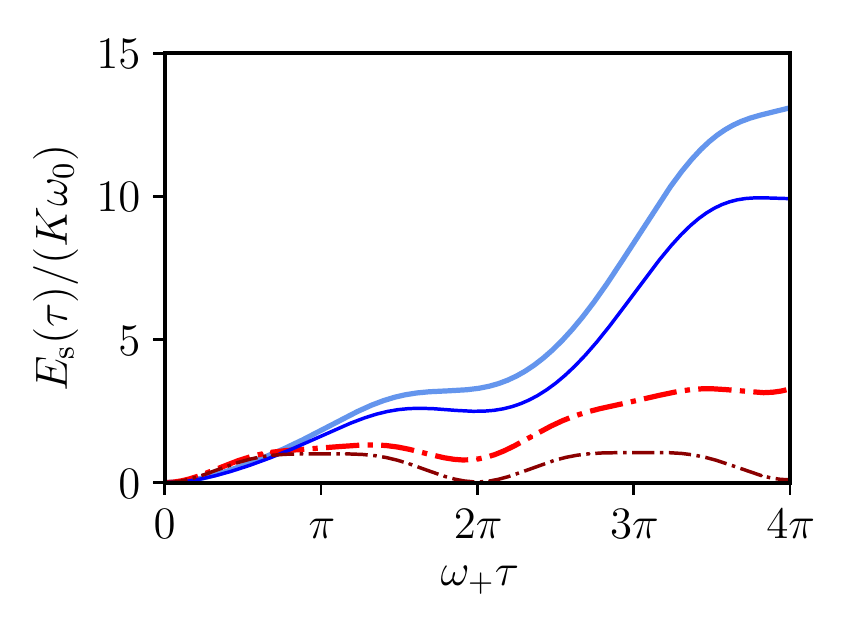}\put(3,66){\normalsize (b)}\end{overpic}\vspace{0.5em} \\
\begin{overpic}[width=0.9\columnwidth]{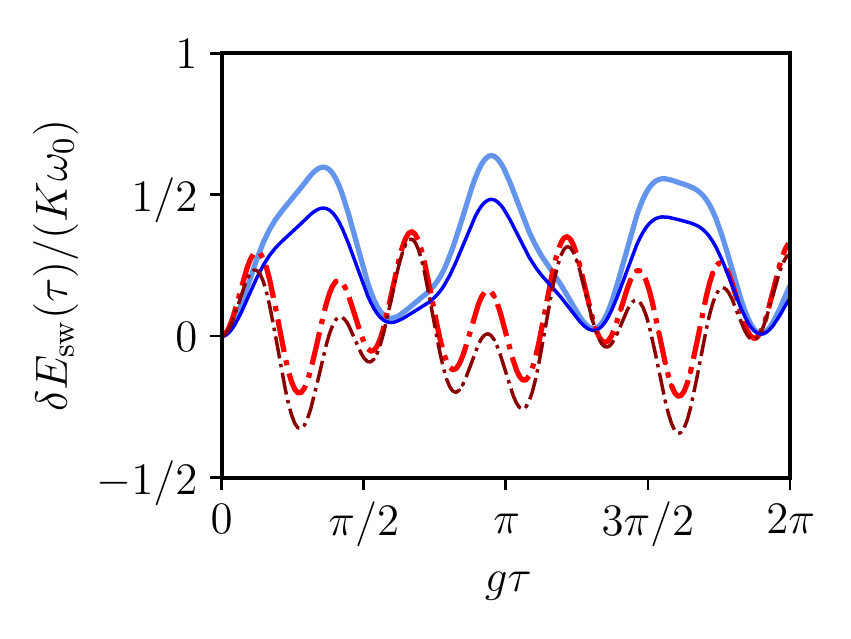}\put(3,66){\normalsize (c)}\end{overpic}\vspace{0.5em} & 
\begin{overpic}[width=0.9\columnwidth]{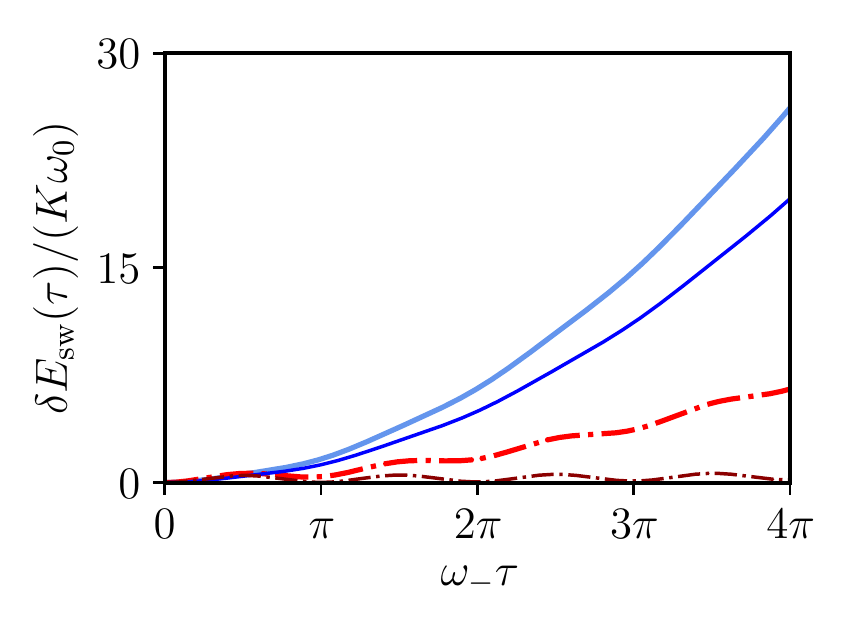}\put(3,66){\normalsize (d)}\end{overpic}\vspace{0.5em}\\
\begin{overpic}[width=0.9\columnwidth]{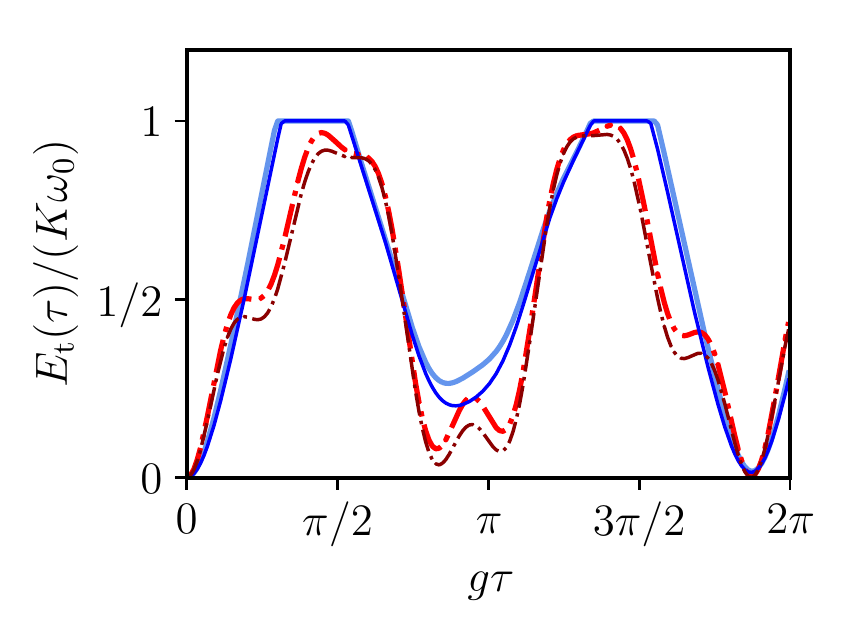}\put(3,66){\normalsize (e)}\end{overpic}\vspace{0.5em} & 
\begin{overpic}[width=0.9\columnwidth]{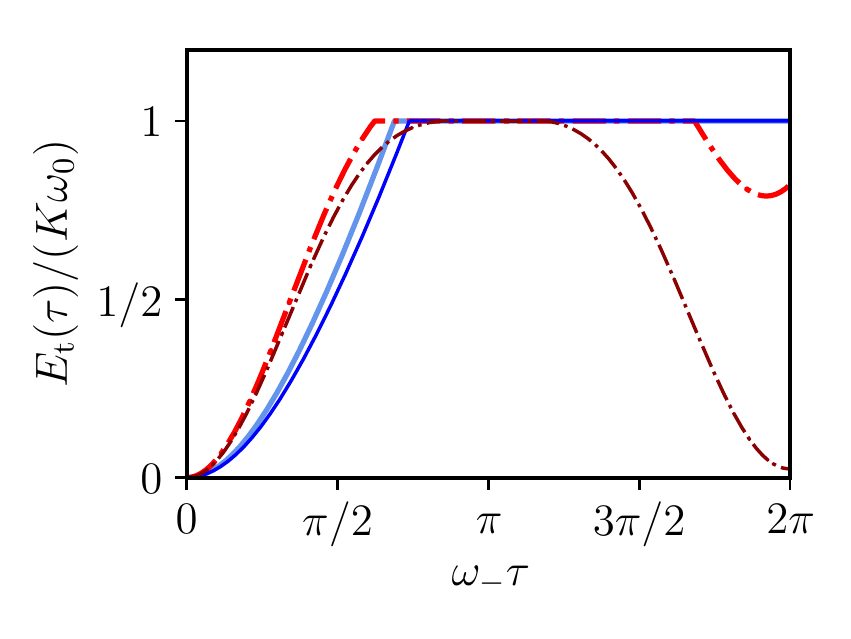}\put(3,66){\normalsize (f)}\end{overpic}\vspace{0.5em}
\end{tabular}
\caption{(Color online)  Panel (a) displays the stored energy $E_{\rm s}(\tau)$ (in units of $K\omega_0$) as a function of $g\tau$, for the case of two coupled QHOs, evaluated by setting $g=0.35~\omega_0$. Different curves refer to results obtained for three different choices of the initial state of the charger: stored energy for an initial Fock or a thermal state evaluated by setting $K=3$ (blue solid line); stored energy for an initial Fock or a thermal state evaluated by setting $K=100$ (dark blue solid line); stored energy for an initial coherent state evaluated by setting $K=3$ (red dash-dotted line); stored energy for an initial coherent state evaluated by setting $K=100$ (dark red dash-dotted line). The same color code is used for all other panels. The stored energy shows oscillations similar to the RWA case (see Fig.~\ref{fig:Eqq}), with counter-rotating terms causing only quantitative corrections. Panel (c) displays the switching energy $\delta E_{\rm sw}(\tau)$ (in units of $K\omega_0$) as a function of $g\tau$, evaluated for $g=0.35~\omega_0$. Panel (e) displays the transferred energy $E_{\rm t}(\tau)$ (in units of $K\omega_0$) as a function of $g\tau$, evaluated for $g=0.35\omega_0$. Panel (b) displays the stored energy $E_{\rm s}(\tau)$ (in units of $K\omega_0$) as a function of $\omega_{+}\tau$ and evaluated for $g\to \omega_0/2$. Due to the vicinity to the critical point, the stored energy increases as a power law. Panel (d) displays the switching energy $\delta E_{\rm sw}(\tau)$ (in units of $K\omega_0$) as a function of $\omega_{-}\tau$, evaluated for $g\to \omega_0/2$. This quantity measures the energy that is externally injected. The power-law increase of this quantity is clear. Panel (f) displays the transferred energy $E_{\rm t }(\tau)$ (in units of $K\omega_0$) as a function of $\omega_{-}\tau$, evaluated for $g\to \omega_0/2$. Only a small amount of the corresponding stored energy seen in panel (b) can be counted as transferred energy, while the majority of the energy is externally injected.\label{fig:Eoo}}
\end{figure*}

We now study the case of two QHOs with counter-rotating terms included in the interaction Hamiltonian, i.e.
\begin{eqnarray}\label{H_ooCRT}
\mathcal{H}_{\rm A}&=&\omega_0 a^\dagger a~, \\ \mathcal{H}_{\rm B}&=&\omega_0 b^\dagger b~, \nonumber \\ \mathcal{H}_1&=&g(a +a^\dagger)(b+ b^\dagger)~. \nonumber
\end{eqnarray} 
In the limit $g \ll \omega_0$ counter-rotating terms, i.e.~terms of the form $a^\dagger b^\dagger$ and $ab$, can be safely neglected~\cite{Schleich_Book} and one recovers Eq.~(\ref{H_oo}).

The full Hamiltonian $\mathcal{H}=\mathcal{H}_{\rm A}+\mathcal{H}_{\rm B}+\mathcal{H}_{1}$, which dictates the dynamical evolution, has eigenvalues $\omega_{\pm}=\sqrt{\omega_0^2\pm2 g\omega_0}$. We therefore assume $|g|/ \omega_0\leq 1/ 2$ in order to guarantee stability of the spectrum.

In order to compute the figures of merit for this beyond-RWA case, it is again useful to use the Heisenberg representation
which in this case yields the following temporal evolutions for the field operators:
 \begin{eqnarray} \label{time_evolved}
a(t)&=&R_{aa}(t)a+R_{ab}(t)b+R_{a a^\dagger}(t)a^\dagger+R_{ab^\dagger}(t)b^\dagger~, \nonumber\\ 
b(t)&=&R_{ba}(t)a+R_{bb}(t)b+R_{ba^\dagger}(t)a^\dagger+R_{bb^\dagger}(t)b^\dagger~,
 \nonumber\\ 
\end{eqnarray}
where the quantities $R_{ij}(t)$ are calculated in Appendix~\ref{appendixRmatrix}. By the same token, the local Hamiltonian for B gets transformed into
 \begin{eqnarray}
&&\mathcal{H}_{\rm B}(t)=\omega_0\big[R^*_{ba}(t)a^\dagger+R^*_{bb}(t)b^\dagger+R^*_{b a^\dagger}(t)a+R^*_{bb^\dagger}(t)b\big] \nonumber\\ && \quad \times \big[R_{ba}(t)a+R_{bb}(t)b+R_{ba^\dagger}(t)a^\dagger+R_{bb^\dagger}(t)b^\dagger\big]~,
\end{eqnarray}
leading to the following expression for the stored energy
 \begin{eqnarray} \label{storedEnRW}
\frac{{E}_{\rm s}(\tau)}{\omega_0}&=&\big[|R_{ba^\dagger}(\tau)|^2+|R_{bb^\dagger}(\tau)|^2\big]\nonumber\\&&+\braket{a^\dagger a}_{\rm A}\big[|R_{ba^\dagger}(\tau)|^2+|R_{ba}(\tau)|^2\big] \nonumber\\ &&+\big[\braket{aa}_{\rm A} R^*_{ba^\dagger}(\tau) R_{ba}(\tau) +{\rm H.c.}\big]~, \end{eqnarray}
where for the sake of simplicity we have denoted the average of an operator $O$ evaluated on the initial state of the charger as $\braket{O}_{\rm A}={\rm tr}_{\rm A}[O \rho_{\rm A}(0)]$. We notice that $\braket{a^\dagger a}_{\rm A}=K$ is the mean value of excitations in the charger at the beginning of the protocol and is proportional to the initial energy, so the first two lines in Eq.~(\ref{storedEnRW}) do not depend on the details of  the initial state. On the contrary, for a coherent state as initial state of A, we have $\braket{aa}_{\rm A}=\alpha^2$, while for both Fock and thermal states of A $\braket{aa}_{\rm A}=0$. Hence, the third line in Eq.~(\ref{storedEnRW}) is different from zero only in the case of a coherent state, while this quantity does not distinguish between a Fock and a thermal state.

The switching energy $\delta E_{\rm sw}(\tau)$ can be calculated as following. We first note that $E_1(0)=0$. We therefore need to calculate only the interaction energy at time $\tau$,  i.e.~$\delta E_{\rm sw}(\tau)=-E_1(\tau)$. With analogous steps to what described just above we find
\begin{widetext}
 \begin{eqnarray} \label{interactionEnRW}
\frac{{\delta E}_{\rm sw}(\tau)}{g}=&-&\Big \{\big[ R_{ab}(\tau)+ R^*_{ab^\dagger}(\tau)  \big]\big[  R_{bb^\dagger}(\tau)+ R^*_{bb}(\tau) \big] +{\rm c.c.} \Big\}  
-\braket{a^\dagger a}_{\rm A} \Big \{\big[ R_{aa}(\tau)+ R^*_{aa^\dagger}(\tau)\big] \big[ R_{ba^\dagger}(\tau)+ R^*_{ba}(\tau) \big] +{\rm c.c.}\Big\} \nonumber \\&- &\Big \{ \braket{aa}_{\rm A}  \big[ R_{aa}(\tau)+ R^*_{aa^\dagger}(\tau)\big]\big[ R_{ba}(\tau)+ R^*_{ba^\dagger}(\tau) \big] +{\rm H.c.}\Big\}~. 
\end{eqnarray}
\end{widetext}
The above considerations for Eq.~(\ref{storedEnRW}) still hold and also Eq.~(\ref{interactionEnRW}) can distinguish only between the coherent state and the other two choices of initial states of the charger. 
It is useful to make a distinction between three situations. In the weak-coupling $|g|/\omega_0\ll 1/2$ regime we can invoke the RWA and apply the analysis described in Sect.~\ref{subsection_oo}. 
The second situation, i.e.~the strong-coupling regime, occurs when $|g|/\omega_0\lesssim1/2$. In this case the counter-rotating terms give quantitative corrections, see Figs.~\ref{fig:Eoo}(a), (c), and (e), while the oscillating behavior of $E_{\rm s}(\tau)$ is still present. 
 Finally, the case $|g|/\omega_0\to1/2$ can be interpreted as a ``critical point'' and the stored energy increases as a power law, see Fig.~\ref{fig:Eoo}(b), (d), and (f). This behavior is due to the fact that one of the two eigenmodes has zero frequency. Indeed, all the observables are functions of the matrix elements $R_{ij}(t)$, which contains the function $\sin(\omega_\pm t)/\omega_{\pm}$. When $\omega_{\pm} \to 0$ we have $\sin(\omega_\pm t)/\omega_{\pm} \to t$, which explains the power-law behavior.

A comment on the strong-coupling and critical regimes is now in order. In the weak-coupling regime, counter-rotating terms can be neglected and rotating terms in ${\cal H}_{1}$ of the form $a b^\dagger + a^\dagger b$ are the ``best interaction Hamiltonian'' from the point of view of energy transfer, since, by definition, they just transfer excitations from A to B and viceversa. On the other hand, in the strong-coupling and critical regimes counter-rotating terms in 
 ${\cal H}_{1}$ of the form $a^\dagger b^\dagger + a b$ cannot be neglected and create/destroy a pair of excitations in the two systems A and B. Now, the impact of these terms is detrimental from the point of view of energy transfer. This is particularly clear in the critical regime, where the both $E_{\rm s}(\tau)$---Fig.~\ref{fig:Eoo}(b)---and the energy of the charger increase as power laws. This growing energy is {\it externally} injected in the system via the time-dependent modulation of the coupling constant $\lambda(t)$ and only a small amount is exchanged between A and B. In summary, in the strong-coupling and critical limits our results cannot be interpreted in terms of pure energy exchange between the two subsystems. The simplest interpretation is, in contrast, in terms of two coupled systems that are externally charged.

\section{Conclusions}
\label{Conclusions}
In this work we presented a systematic classification and analysis of several simplified models of 
energy transfer  for quantum batteries and of their associated charging processes.  Our approach, based only on different combinations of two-level systems and harmonic oscillators, allowed us to derive exact results without the necessity of introducing any particular assumption or approximation.
The set of models considered in this work covers many paradigmatic situations including the non trivial one  when the interaction does not preserve the total number of excitations.

Some of the results obtained in this work for toy models of quantum batteries 
are expected to hold in general. 
For example, the scaling of the charging time with an inverse power-law of the charger energy is expected to be general---see also Ref.~\onlinecite{Ferraro17}. Moreover, we believe that the fact that quantum coherences in the basis of the eigenstates of the charger Hamiltonian are not a necessary ingredient in order to achieve optimal figures of merit is a general result, provided that no counter-rotating terms are at play. Finally, the fact that the strong-coupling regime is not suitable for studying the ideal scenario of pure energy exchange between charger and battery is also expected to hold true in more complicated models.

Possible future outlooks and applications of our work could be: theoretical or experimental implementations of our models in specific systems and real devices, the development of a more detailed analysis taking into account also the presence of energy fluctuations, the extension of the considered models to systems of arbitrary dimension, the presence of loss or other noisy mechanisms, and charging of the battery via an external classical field~\cite{farina_arxiv_2018}. 
Since it would be highly desirable for our quantum battery to store energy for a relatively long time, a thorough study of the role of dissipative effects during the storage step should also be carried out and is left for future work.

Within the general context of quantum enhanced technologies, we hope that the simple yet exactly solvable models of quantum batteries considered in this work could represent a solid starting point stimulating new ideas and further research lines. 

\acknowledgments
We gratefully thank M. Campisi, F.M.D. Pellegrino, D. Ferraro, P.A. Erdman, V. Cavina, and M. Keck for useful discussions.

\appendix

\section{Details on the calculation of Eq.~(\ref{time_evolved})}
\label{appendixRmatrix}
In this Appendix we show the details of the calculation of Eq.~(\ref{time_evolved}). First of all, in order to find the time evolution of the ladder operator it is useful to diagonalize the problem. We define as $A$ the vector made by the ladder operators involved in the problem:
 \begin{eqnarray} 
 A=\begin{pmatrix}
 a\\
 b\\
 a^\dagger\\
 b^\dagger
 \end{pmatrix}~.
\end{eqnarray}
In a similar way we denote as $\gamma$ the vector of made by the operators that diagonalize the Hamiltonian, i.e.
 \begin{eqnarray} 
 \gamma=\begin{pmatrix}
 \gamma_-\\
 \gamma_+\\
 \gamma^\dagger_-\\
 \gamma^\dagger_+
 \end{pmatrix}~.
\end{eqnarray}
Diagonalization of the Hamiltonian consists in finding the transformation $A=M\gamma$, where $M$ is a $4\times 4$ matrix. Finding $M$ is a straightforward textbook task~\cite{Schleich_Book}. In the Heisenberg representation the vector evolved at time $t$, $\gamma(t)$, is related via a diagonal matrix $D$ to the vector eingenmodes at the initial time $\gamma(t)=D\gamma$:
\begin{eqnarray} 
D=\begin{pmatrix}
 e^{-i\omega_-t} &0 &0&0\\
0 &e^{-i\omega_+t} &0&0\\
 0 &0& e^{i\omega_-t} &0\\
0 &0&0& e^{i\omega_+t} 
 \end{pmatrix}~.
\end{eqnarray}
Our goal is to find $A(t)=R (t)A$, where $R(t)$ is the matrix in Eq.~(\ref{time_evolved}). In order to find such transformation,  we express $A$ in terms of the eigenmodes $\gamma$, we evolve the eigenmodes, and then we express the eingemodes in terms of the initial ladder operators, using the inverse transformation $M^{-1}$, i.e. $A(t)=\big[M ~D(t)~M^{-1}\big]A$. Hence we find:
 \begin{eqnarray} \label{Rmatrix1} 
R_{ aa}(t)&=&\frac{1}{2} \Big[ \cos(\omega_-t)+\cos(\omega_+t) \Big]\nonumber \\
 & -&\frac{i}{2}  \Big[\big(\frac{\omega_0^2+\omega_-^2}{2\omega_0}\big)\frac{\sin(\omega_-t)}{\omega_-}+\big(\frac{\omega_0^2+\omega_+^2}{2\omega_0}\big)\frac{\sin(\omega_+t)}{\omega_+} \Big]~,\nonumber\\   
 \\
 R_{ ab}(t)&=&\frac{1}{2} \Big[- \cos(\omega_-t)+\cos(\omega_+t) \Big]\nonumber\\\nonumber
  & -&\frac{i}{2}  \Big[-\big(\frac{\omega_0^2+\omega_-^2}{2\omega_0}\big)\frac{\sin(\omega_-t)}{\omega_-}+\big(\frac{\omega_0^2+\omega_+^2}{2\omega_0}\big)\frac{\sin(\omega_+t)}{\omega_+} \Big]~,\nonumber \\   
    R_{ aa^\dagger}(t)&=&\frac{ig}{2} \Big[ \frac{   \sin (\omega_{-} t)}{ \omega_{-} }-\frac{ 
   \sin (\omega_{+} t))}{ \omega_{+}} \Big]~,  \nonumber \\
  R_{ ab^\dagger}(t)&=&-\frac{ig}{2}\Big[ \frac{   \sin (\omega_{-} t)}{ \omega_{-} }+\frac{ 
   \sin (\omega_{+} t))}{ \omega_{+}} \Big] ~.\nonumber
\end{eqnarray}
From the fact that the Hamiltonian is symmetric with respect to the exchange $a\leftrightarrow b$, we have:
 \begin{eqnarray} \label{Rmatrix2}
R_{ ba}(t)&=&R_{ ab}(t)~,
 R_{ bb}(t)= R_{ aa}(t)~,\\
 R_{ ba^\dagger}(t)&=& R_{ ab^\dagger}(t)~,  
    R_{ bb^\dagger}(t)=R_{ aa^\dagger}(t)~. \nonumber
\end{eqnarray}

\end{document}